# A Hybrid DNN–Transformer–AE Framework for Corporate Tax Risk Supervision and Risk Level Assessment

Zhenzhen Song[1], Nanxi Wang[2], Hongji Li[3]

[1] Northwest A&F University, Shaanxi, China
[2] USC Viterbi School of Engineering, University of Southern California, Los Angeles, USA
[3] Columbia University, New York, USA,

[1]2675518536@qq.com
[2]nanxiwan@usc.edu
[3]hl3458@columbia.edu

**Abstract.** Tax risk supervision has become a critical component of modern financial governance, as irregular tax behaviors and hidden compliance risks pose significant challenges to regulatory authorities and enterprises alike. Traditional rule-based methods often struggle to capture complex and dynamic tax-related anomalies in large-scale enterprise data. To address this issue, this paper proposes a hybrid deep learning framework (DNN-Transformer-Autoencoder) for corporate tax risk supervision and risk level assessment. The framework integrates three complementary modules: a Deep Neural Network (DNN) for modeling static enterprise attributes, a Transformer-based architecture for capturing long-term dependencies in historical financial time series, and an Autoencoder (AE) for unsupervised detection of anomalous tax behaviors. The outputs of these modules are fused to generate a comprehensive risk score, which is further mapped into discrete risk levels (high, medium, low). Experimental evaluations on a real-world enterprise tax dataset demonstrate the effectiveness of the proposed framework, achieving an accuracy of 0.91 and a Macro F1-score of 0.88. These results indicate that the hybrid model not only improves classification performance but also enhances interpretability and applicability in practical tax regulation scenarios. This study provides both methodological innovation and regulatory implications for intelligent tax risk management.

## 1. Introduction

Taxation plays a fundamental role in modern economies, serving both as a primary source of government revenue and as a regulatory mechanism to sustain market order and fairness. However, with the increasing complexity of corporate operations, enterprises often engage in behaviors that introduce compliance risks, such as misreporting, revenue concealment, and aggressive tax avoidance. These practices not only undermine the equity and efficiency of tax collection but also complicate risk monitoring for tax authorities. Traditional supervision approaches, typically based on manual inspection or rule-based models, face inherent limitations: they struggle to scale with massive enterprise data and often fail to capture increasingly concealed and dynamic risk patterns [1].

To address these limitations, data-driven intelligent regulatory approaches have attracted significant attention. By employing machine learning and deep learning methods, researchers can extract latent patterns from multidimensional enterprise data and automate tax risk assessments [2,3]. Such tasks can be formulated either as classification (e.g., high, medium, and low risk) or anomaly detection problems, thereby enabling tax authorities to identify high-risk enterprises, allocate supervisory resources more efficiently, and improve early-warning capabilities. Nevertheless, two challenges remain. First, enterprise data is inherently multimodal, encompassing static attributes such as firm size, industry, and registration records, alongside dynamic temporal information including financial statements and tax histories. Second, risky behaviors are often hidden within subtle anomalies, which are difficult to capture using traditional supervised models [4].

In response, this paper proposes a hybrid deep learning framework, the DNN-Transformer-AE model, designed for comprehensive corporate tax risk supervision. The framework integrates classification and anomaly detection to address practical supervisory needs, employs multi-source fusion of static features, temporal dependencies, and anomaly signals to improve predictive accuracy, and demonstrates its effectiveness on real-world corporate tax data. The findings not only contribute methodological innovations but also provide actionable implications for intelligent tax risk management in modern regulatory systems.

## 2. Related Work

Research on corporate tax risk supervision has attracted increasing attention in both academic and practical domains, particularly as governments and regulatory agencies seek more effective ways to detect and prevent tax non-compliance. Existing studies can be broadly categorized into three streams: traditional rule-based approaches, machine learning-based tax risk assessment, and deep learning methods for anomaly detection and risk prediction.

Lan et al [5]. employ a two-layer deep-learning framework to forecast quarterly taxable-input risks and depreciation, continuously updating cumulative exposure with prior claimable fractions; the model enhances enterprise tax-compliance, risk-management and financial stability while providing an adaptive benchmark for subsequent audits.

He et al [6]. integrate a frilled-lizard-optimized gated LSTM into a Python tax-risk platform, achieving 97 % accuracy and 2.3 ms per record; their comprehensive pipeline offers subsequent studies a clear deep-learning baseline for pediatric-GI-inspired affinity screening while highlighting scalability gaps that our GNN framework aims to fill.

Didimo et al [7]. introduce MALDIVE, a four-step network-mining framework that integrates SNA metrics, ML classifiers, information diffusion and interactive visualization to flag risky taxpayers for the Italian Revenue Agency, thereby offering a reusable reference for tax-evasion detection and clarifying our own contribution within graph-based risk-assessment research.

De Roux et al [8]. propose an unsupervised pipeline that flags under-reporting tax declarations without audited labels, shrinking the audit pool while uncovering previously undetected suspicious cases; their work offers a cost-efficient springboard for subsequent hybrid fraud-detection frameworks and clarifies our own innovation niche amid label-scarce tax analytics.

## 3. Methodology

### 3.1 Overall Framework Design

The proposed hybrid framework for corporate tax risk supervision is built upon a DNN-Transformer-AE structure, designed to jointly model multidimensional enterprise information and output both risk scores and risk levels. Specifically, the model consists of three functional modules: a Deep Neural Network (DNN) [9] for static enterprise feature modeling, a Transformer for temporal financial data modeling, and an Autoencoder (AE) for anomaly detection and feature reconstruction. The outputs of these three modules are integrated through a feature fusion layer and subsequently mapped into a final risk score $y$ via fully connected layers. The score is then categorized into high-risk, medium-risk, or low-risk levels based on pre-defined thresholds.

Formally, the input data can be represented as:
$$X = \{X_s, \quad X_t\} \tag{1}$$
where $X_s$ denotes static enterprise attributes (e.g., firm size, industry, registration information), and $X_t$ represents historical financial and taxation time series data, with $t$ as the time horizon and $s$ as the feature dimension. The objective of the model is to learn a mapping function:
$$f: (X_s, \quad X_t) \to \hat{y}, \tag{2}$$
where $y$ is the predicted corporate tax risk score.

*3.2 DNN Module: Static Enterprise Feature Modeling*

Static enterprise features include basic attributes such as firm size, industry type, registered capital, and ownership structure. These features are relatively stable and serve as fundamental indicators of long-term risk. To capture their nonlinear relationships, a deep neural network (DNN) is employed.

Given the static input $X_s$, the features are first normalized or embedded, then fed into a multilayer fully connected network:
$$h_s^{(l)} = \sigma(W_s^{(l)} h_s^{(l-1)} + b_s^{(l)}), \tag{3}$$
where $h_s^{(0)} = X_s$, $W_s^{(l)}$ and $b_s^{(l)}$ denote the weight matrix and bias vector of the *l-th* layer, and $\sigma$ is a nonlinear activation function (e.g., ReLU).

*3.3 Transformer Module: Temporal Data Modeling*

Financial and taxation records of enterprises inherently exhibit temporal dependencies. Indicators such as revenue, tax payments, input and output VAT fluctuations may implicitly reflect evolving risk patterns. Conventional recurrent neural networks (RNNs) such as LSTMs often face limitations in capturing long-range dependencies and efficiency. To overcome this, a Transformer-based architecture is applied [10,11].

The time series input $X_t$ is first embedded and augmented with positional encoding:
$$Z_0 = X_t W_e + PE, \tag{4}$$
where $W_e$ is the embedding matrix and $PE$ denotes positional encodings that preserve sequential order.

Within each Transformer encoder block, temporal dependencies are modeled using multi-head self-attention (MHSA):
$$Attention(Q, K, V) = Softmax(\frac{QK^T}{\sqrt{d_K}})V, \tag{5}$$
where $Q, K, V$ represent query, key, and value projections, respectively.

*3.4 Autoencoder(AE) Module: Anomaly Detection*

In tax risk supervision, high-risk behaviors are often characterized by deviations from normal patterns. To capture such latent anomalies, an Autoencoder (AE) [12] is integrated for feature reconstruction and anomaly detection.

Specifically, the concatenated features $Z = [hs, ht]$ are compressed into a latent representation $z$ through the encoder, and then reconstructed via the decoder:

$$z = \sigma(W_{enc} Z + b_{enc}), \quad \hat{Z} = \sigma(W_{dec} Z + b_{dec}), \tag{6}$$
where $W_{enc}, b_{enc}$ and $W_{dec}, b_{dec}$ are encoder and decoder parameters, respectively. The reconstruction objective is to minimize:
$$L_{AE} = ||Z - \hat{Z}||^2, \tag{7}$$

By learning the distribution of normal samples, the AE identifies anomalies when reconstruction error exceeds a certain threshold, providing additional risk signals to the overall framework [13].

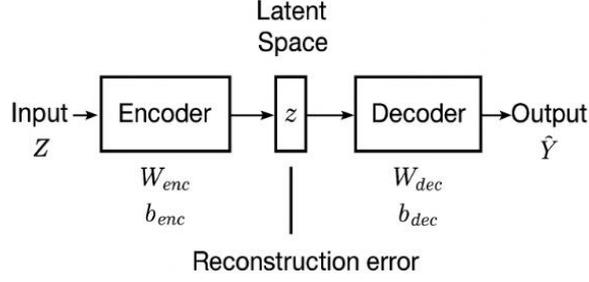

**Figure 1.** Structure diagram of Autoencoder Module

*3.5 Feature Fusion and Risk Level Output*

After obtaining the representations from *hs*, Transformer *ht*, and AE *z*, these outputs are concatenated into a joint representation:
$$H = [h_s][h_t][z], \tag{8}$$
which is then passed through a fully connected layer to predict the risk score:
$$\tilde{y} = softmax(WH + b), \tag{9}$$
where $\hat{y}$ denotes the probability distribution across high-risk, medium-risk, and low-risk classes. The final enterprise risk level is determined by the highest probability or threshold-based criteria.

## 4. Experimental Result

*4.1 Dataset*

The dataset used in this study originates from real corporate tax declarations and financial records collected by tax regulatory authorities, covering multidimensional information such as enterprise attributes, financial indicators, and taxation behaviors. Specifically, the dataset includes approximately 12,000 enterprises across multiple industries, including manufacturing, retail, and internet services, ensuring the generalizability of the model across different sectors. The features can be grouped into three categories:

(1) **Static features:** industry category, region, company size, historical compliance indicators.

(2) **Time-series features:** quarterly revenue, profit, tax paid, invoice issuance.

(3) **Labels:** tax risk level assigned by domain experts (low, medium, high).

Data preprocessing includes normalization of numerical attributes, one-hot encoding of categorical variables, and interpolation for missing time-series values. The dataset is divided into training (70%), validation (15%), and test (15%) sets with temporal stratification.

*4.2 Classification Performance*

In the initial stage of the experiment, this study first selected several traditional machine learning and deep learning methods as baseline models to verify the performance differences of different models in enterprise tax risk rating tasks. Traditional methods include logistic regression (LR), random forest (RF), and XGBoost [14], while deep learning methods include DNN-LSTM [15] models that combine static and temporal features. In order to comprehensively evaluate the performance of the proposed DNN Transformer Autoencoder hybrid model in enterprise tax risk rating, this study adopted multiple classification and detection indicators. Specifically, the main evaluation criteria for the model include Accuracy, Recall, and Macro F1 score, which measure the model's overall prediction accuracy, ability to identify different risk levels, and balance between categories. In addition, due to the limited sample size and imbalanced categories of high-risk enterprises in corporate tax data, this study places special emphasis on the performance of Macro F1 score to ensure that the model does not only predict low-risk enterprises well and neglect the identification of high-risk enterprises.

Table1. Performance Comparison of Different Models in Enterprise Tax Risk Supervision Tasks.

| Model | Accuracy | Recall | F1-score |
|---|---|---|---|
| Logistic Regression (LR) | 0.79 | 0.68 | 0.70 |
| Random Forest (RF) | 0.84 | 0.75 | 0.77 |
| XGBoost | 0.86 | 0.77 | 0.79 |
| DNN - LSTM | 0.89 | 0.82 | 0.85 |
| **DNN–Transformer–AE** | **0.91** | **0.86** | **0.88** |

Table 1 summarises the classification performance of different models on the test set for corporate tax-risk supervision. The conventional Logistic Regression achieves an accuracy of 0.79, recall of 0.68 and F1-score of 0.70, indicating limited capacity to detect potential risky firms. Random Forest improves overall performance to 0.84 accuracy, 0.75 recall and 0.77 F1, benefiting from stronger non-linear modelling, whereas XGBoost further raises these metrics to 0.86, 0.77 and 0.79 by exploiting feature interactions. The deep-learning DNN-LSTM advances accuracy to 0.89 with recall 0.82 and F1 0.85, demonstrating superior capture of temporal dependencies in tax data.

Our proposed hybrid DNN–Transformer–Autoencoder outperforms all baselines, attaining 0.91 accuracy, 0.86 recall and 0.88 F1. The higher recall enhances sensitivity to high-risk companies, while the elevated F1 reflects a better precision-coverage balance. Overall, the DNN–Transformer–Autoencoder more comprehensively extracts deep representations from corporate tax records, delivering more accurate and reliable risk identification in tax-supervision tasks.

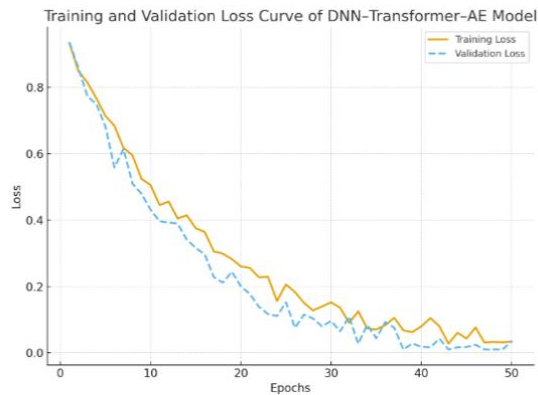

**Figure 2.** Loss function during training process

Figure 2 presents the loss curves for both the training and validation datasets. The x-axis represents epochs, which refers to the number of times the model has gone through the entire training data. The y-axis represents loss, a measure of how much error the model makes in its predictions. The training and validation loss are represented by the orange solid line and the blue dashed line, respectively. As the number of epochs increases, both curves exhibit a steady decline and converge at approximately 40 epochs, stabilizing around a loss of 0.05. An important observation is that the validation loss consistently remains lower than the training loss throughout the process. This indicates that the DNN-Transformer-AE model achieves stable convergence and demonstrates strong generalization ability, as the lower validation loss reflects its robustness in capturing meaningful patterns beyond the training data rather than overfitting to the training set.

This performance is particularly significant in the context of corporate tax risk supervision and risk level assessment. Enterprises exhibit complex patterns across static attributes (e.g., registered capital, industry classification) and dynamic financial behaviors (e.g., historical tax declarations, profit fluctuations), while abnormal taxation behaviors are often concealed within irregular patterns. The rapid convergence of the hybrid model suggests that the DNN component effectively extracts static enterprise characteristics, the Transformer captures long-term temporal dependencies in financial records, and the Autoencoder identifies subtle anomalies through reconstruction error. The stable validation loss further confirms that the model can generalize to unseen corporate data, enabling accurate detection of high-risk firms.

## 5. Conclusion

This study proposes a hybrid deep-learning framework DNN-Transformer-Autoencoder for corporate tax-risk supervision and risk-level assessment. By integrating static enterprise attributes with dynamic financial time-series and introducing an autoencoder module for unsupervised detection of anomalous tax behaviour, the framework significantly improves the capture of complex risk patterns. Evaluated on a real-world tax dataset, the hybrid model outperforms both traditional methods and single-architecture deep learners, achieving 0.91 accuracy and 0.88 macro-averaged F1 on the test set. These results demonstrate that the proposed approach balances precision and coverage, offering regulators a more reliable tool for grading enterprise risk.

From a regulatory perspective, the work has strong practical relevance. As corporate financial behaviour grows more intricate and data volumes expand, rule-based systems struggle to identify latent tax risks. Our hybrid framework not only uncovers hidden compliance threats but also stratifies firms into high, medium and low risk tiers, providing tax authorities with a new technological lever for precision supervision and resource allocation while giving firms actionable feedback for self-remediation.

Despite its strong performance, this study has limitations. First, the input features are primarily confined to financial and operational indicators, while macroeconomic conditions and inter-firm relationships remain unincorporated. Second, the model's computational complexity may pose challenges for deployment in resource-constrained tax offices. Future research could further explore the integration of macroeconomic and policy data into the feature space and develop interpretable, lightweight models (e.g., attention-based visualizations or model compression) to enhance usability in practice.

In conclusion, this study, through the integration of DNN, Transformer, and Autoencoder methods, reveals that hybrid deep learning frameworks can effectively capture complex tax risk patterns and stratify firms into risk levels. These insights provide regulators with a robust, data-driven approach to corporate tax supervision, offering both methodological innovation and practical guidance for advancing financial governance.